\documentclass{article}
\usepackage{latexsym}

\def\cH{{\cal H}} \def\cP{{\cal P}}

\def\ket#1{\mid~\!\!\!{#1}~\!\!\rangle}
\def\bra#1{\langle~\!\!{#1}~\!\!\!\mid}
\def\IF{if and only if }
\def\QM{quantum mechanics }
\def\qm{quantum mechanics}

\def\${\enskip$}
\def\M{measurement }
\def\m{measurement}
\def\Q{quantum }

\def\D{Definition }

\def\P{Proposition }

\def\C{Corollary }
\def\R{Remark }
\def\T{Theorem }

\begin{document}

{\bf \large \noindent Delayed Twin Observables\\ Are They a Fundamental Concept in Quantum Mechanics?\\

\begin{quote}

\normalsize \rm Fedor Herbut(E-mail:
fedorh@sanu.ac.rs)\\

{\it \footnotesize \noindent Serbian
Academy of Sciences and
Arts, Knez Mihajlova 35,\\
11000 Belgrade, Serbia}\\

\normalsize \noindent Opposite-subsystem twin events and
twin observables, studied previously in the context of
distant correlations,  are first generalized to pure states
of not-necessarily-composite systems, and afterwards they
are further generalized to delayed twins that are due to  unitary
evolution of the \Q system. The versatile aspects of delayed twin observables are
studied in terms of necessary and sufficient conditions to make possible various applications. Three of these are sketched: Preparation of some \Q experiments, easy solution of a puzzle in an important Scully et al. real experiment, and, finally, it is shown that exact \M in \QM is an example of opposite-subsystem delayed twins in bipartite pure states.

\end{quote}

\rm
\section{INTRODUCTION}

Opposite-subsystem twin observables in bipartite systems
were previously studied in a number of articles
\cite{FHMVAnnPhys}, \cite{MVFHJMP},  \cite{FHQLOG}, \cite{FHDZ}.
To illustrate subsystem twins,  we consider the well known
singlet state
$$\ket{\Psi}_{AB} \equiv (1/2)^{1/2}\Big(\ket{+}_A\ket{-}_B-
\ket{-}_A\ket{+}_B\Big).$$ The plus and minus stand for
the spin-up and spin-down respectively along the z-axis.
The twin observables (operators) in this bipartite state are
the z-projections: \$s_A^z\$ and \$-s_B^z\$. They are twins
because they act equally on the composite state vector \$\ket{\Psi}_{AB}\$ (cf Definitions 6 and 2 below).

The singlet state vector is written in the form of a
twin-adapted canonical Schmidt (or bi-orthogonal) expansion.
"Twin-adapted" means that the subsystem state vectors
appearing in the expansion are eigen-vectors of the twin
observables; "canonical" means that the squares of the expansion coefficients are the eigenvalues of the reduced density operators (subsystem states). The mentioned properties of the expansion of the singlet state vector are general in the sense that every state vector of a bipartite system can be written as a twin-adapted canonical Schmidt expansion (cf section 2 in \cite{FHZurek}).

The most important properties of twin observables are that they have the same probabilities in the state at issue, and
that they give, by ideal \m , the same change of the composite-system state.

In this article the concept of twin observables is first
generalized to systems that are not necessarily composite.
Events (projectors), as the simplest observables, are studied in detail, and the results  are utilized for general observables (but restricted to those that do not have a continuous part in their spectrum). The main purpose of this generalization is that it is the basis for further generalization.

The next generalization is to delayed twins, in which
a unitary evolution operator is allowed to separate the
twin observables.

The last section is devoted to three applications.\\

\section{FIRST GENERALIZATION:\\ TWIN EVENTS IN PURE  STATES OF\\ NOT-NECESSARILY-COMPOSITE SYSTEMS}

The terms 'event' and 'projector',  'observable' and 'Hermitian operator', as well as 'pure state' and state vector (vector of norm one) are used interchangeably throughout. If \$E\$ denotes an event, then \$E^c\$ denotes the opposite event (the ortho-complementary or simply complementary projector) \$E^c\equiv I-E\$, where \$I\$ is the identity operator. The reader should be reminded that the probability of an event \$E\$ in a state \$\ket{\psi}\$ is given by the
expression \$\bra{\psi}E\ket{\psi}\$ in \qm .

We are going to need a known but perhaps not well known lemma.\\

{\bf Lemma.} An event (projector) \$P\$ is {\bf certain} (has probability one) in a pure state \$\ket{\phi}\$ \IF \$\ket{\phi}=P\ket{\phi}\$ is valid.\\

{\bf Proof.} The claim is almost obvious if one has in mind the identities \$\ket{\phi}=P\ket{\phi}+P^c\ket{\phi}\$ and
\$1=||\ket{\phi}||^2= ||P\ket{\phi}||^2+||P^c\ket{\phi}||^2\$.\hfill $\Box$\\

We will also need the following simple fact. If \$E\$ is an event and \$\ket{\psi}\$ a state vector, then the latter can be viewed as consisting of two {\bf coherent} (cf the Introduction in \cite{FHcoher}) component state vectors, each having a sharp value 1 or 0 of the observable \$E=1E+0E^c\$: $$\ket{\psi}=
||E\ket{\psi}||\times \Big(E\ket{\psi}\Big/||E\ket{\psi}||
\Big)+
||E^c\ket{\psi}||\times \Big(E^c\ket{\psi}\Big/||E^c\ket{\psi}||\Big).\eqno{(1)}$$

{\bf Definition 1.}  If an event \$E\$ has, as an observable, the sharp value 1 in a state \$\ket{\psi}\$, then we say that \$E\$ 'has occurred' in the latter; if \$E\$ has the sharp value 0, then the opposite event \$E^c\$ has occurred in it. If a state \$\ket{\psi}\$ changes into its first component in decomposition (1), a state in which \$E\$ has occurred, then we say that \$E\$ {\bf occurs ideally}  or 'in an ideal way' or 'in ideal \m ' in \$\ket{\psi}\$. (We are dealing with the selective, pure-state L\"uders change-of-state formula \cite{Lud}, \cite{Messiah}, \cite{Laloe}).\\

{\bf Definition 2.} Let \$E\$ and \$F\$ be two events and \$\ket{\psi}\$ a state vector. If the projectors act equally on the vector
$$E\ket{\psi}=F\ket{\psi},
\eqno{(2)}$$ then we have {\bf twin events} or events that are twins.\\

The following theorems  present twin events in a versatility of mathematical forms and {\bf physical meanings}.

{\bf Theorem 1.} Two (simple and composite) necessary and sufficient conditions - alternative definitions - (i) and (ii) for two events \$E\$ and \$F\$ to be {\bf twins} in a pure state \$\ket{\psi}\$ read:

{\bf (i)} The opposite events are twins in \$\ket{\psi}\$: $$ E^c\ket{\psi}=F^c\ket{\psi}.\eqno{(3)}$$

{\bf (ii) (a)} The events \$E\$ and \$F\$ have the {\bf same} {\bf probability} in \$\ket{\psi}\$: $$ \bra{\psi}E\ket{\psi}=\bra{\psi}F\ket{\psi}.
\eqno{(4a)}$$

{\bf (ii) (b)} If the probability (given by (4a)) is positive, then the two events bring about the {\bf same change of the state} {\bf in ideal occurrence}, i. e., \$\ket{\psi}\$  changes into:
$$E\ket{\psi}\Big/||E\ket{\psi}||=F\ket{\psi}\Big/
||F\ket{\psi}||.\eqno{(4b)}$$\\

{\bf Proof.} {\bf (i)} Subtracting (2) from \$I\ket{\psi}=I\ket{\psi}\$, one obtains condition (3). Since (2) and (3) are in a symmetrical relation, also the sufficiency of (3) for (2) is proved.

{\bf (ii)} Condition (ii) (a) follows obviously from (2), and, if the probabilities in (4a) are positive, so does (ii) (b). Conversely, if (ii) (a) is valid and the probabilities in (4a) are positive, then the denominators in (ii) (b) are equal, then (ii) (b), if valid, it makes also the nominators equal. Thus (2) is valid.

If, on the other hand, the probabilities in (4a) are zero, which amounts to \$||E\ket{\psi}||=||F\ket{\psi}||=0\$, then (2) is obviously implied in the form \$E\ket{\psi}= F\ket{\psi}=0\$ .\hfill $\Box$\\

The conditions (i) and (ii) in Theorem 1 cover both extremes: \$0=E\ket{\psi}=
F\ket{\psi}\$ and \$E\ket{\psi}=
F\ket{\psi}=\ket{\psi}\$ of twin projectors. In many concrete cases the first extreme is irrelevant. Now we exclude it, and give two more conditions.\\

{\bf Theorem 2.} Let \$E\$ and \$F\$ be two events and \$\ket{\psi}\$ a pure state such that both events have positive probability in it. Then, two (composite) necessary and sufficient conditions - alternative definitions - (i) and (ii) for twin events are:

{\bf (i) (a)}  \$F\$ is a certain event in the state \$E\ket{\psi}\Big/||
E\ket{\psi}||\$ (which is the first component in (1)): \$E\ket{\psi}\Big/||E\ket{\psi}||=F
\Big(E\ket{\psi}\Big/||E\ket{\psi}||\Big)\$.

{\bf (i) (b)} {\it Vice versa}, \$E\$ is a certain event in the state \$F\ket{\psi}\Big/||F\ket{\psi}||\$: \$F\ket{\psi}\Big/||F\ket{\psi}||=E\Big(
F\ket{\psi}\Big/||F\ket{\psi}||\Big)\$.
This state is a component in the coherent state decomposition $$\ket{\psi}=||F\ket{\psi}||\times\Big( F\ket{\psi}\Big/
||F\ket{\psi}||\Big)+||F^c\ket{\psi}||\times\Big( F^c\ket{\psi}\Big/||F^c\ket{\psi}||\Big),\eqno{(5)}$$
Finally,

{\bf (i) (c)} the projectors \$E\$ and \$F\$ commute on \$\ket{\psi}\$: \$EF\ket{\psi}=FE\ket{\psi}\$.\\

{\bf (ii) (a)} Claim (i) (a) is valid.

{\bf (ii) (b)} The  analogous claim is valid for the opposite
events \$E^c\$ and \$F^c\$: \$F^c\$ is a certain event in the state \$E^c\ket{\psi}\Big/||E^c\ket{\psi}||\$,
which is the second component in the coherent state decomposition (1).\\

{\bf Proof.\bf (i)}  On account of idempotency of projectors, (2) implies conditions (i) (a), (b), (c).  Conversely, the latter three evidently imply (2).

{\bf (ii)} That (2) has (ii) (a) as its consequence has already been proved
in the preceding passage, and (2) implies (ii) (b) due to the equivalence of (2) with condition (3) (cf Theorem 1 (i)). Conversely if (ii) (b) is satisfied, then
\$(I-E)\ket{\psi}=(I-F)(I-E)\ket{\psi}\$ is valid
(cf the Lemma). This gives \$0=(-F+FE)\ket{\psi}\$.
Substituting here \$FE\ket{\psi}=E\ket{\psi}\$, which is
implied by (ii) (a), one obtains \$(-F+E)\ket{\psi}=0\$,
i. e., we are back at (2) as claimed.
\hfill $\Box$\\

{\bf Remark 1.} One should note that definition (2) and conditions (i) and (ii) of Theorem 1 and condition (i) of Theorem 2 for twin events are {\bf symmetric} in
the events, whereas definition (ii) in Theorem 2 is {\bf asymmetric}
in the events, but it is symmetric in taking the opposite event.
Besides, definitions (2) and (3) are mathematical,
in particular algebraic, whereas conditions (ii) of Theorem 1 and conditions (i) and (ii) of Theorem 2 are expressed in terms of entities with obvious physical meaning (with the exception of condition (i) (c) of Theorem 2, which is algebraic).\\

In the mentioned previous studies of twin events (and twin observables, cf the Introduction) only
the algebraic definition (2) was used, and the concept was confined to composite systems. In this article we will not explore where the equivalent properties given in Theorems 1 and 2 find application. We confine ourselves to their utilization in the investigation of delayed twins that are going to be introduced in the next section.\\

{\bf Remark 2.} It is evident in Theorems 1 and 2 that all claims are made in terms of the eigen-projectors \$E\$ and \$E^c\$ of the binary observable \$E=1E+0E^c\$; the eigenvalues play no role. Hence, one can take an arbitrary {\bf binary observable} \$O=o_1E+o_2E^c,\enskip o_1\not= o_2\$, and Theorems 1 and 2 are valid for it. (In \C 1 below we go to more general observables.)\\

The projectors \$E\$ and \$F\$ are not required to be
necessarily distinct. However trivial is  the claim that
'each projector is its own twin', allowing for this makes
relation (2) a reflexive, symmetric and transitive one,
i. e.. an  equivalence relation in the set \$\cP(\cH)\$ of all projectors
in the state space (Hilbert space) \$\cH\$
of the \Q system. One may wonder if these {\bf equivalence
classes} have a simple structure.
An affirmative answer was given in the more general case of
mixed-or-pure states (density operators) in a recent study
\cite{FHAMP} ( see section 5 and Theorem 5.3. there). The classes are denoted by  \$[E]\$; it is the class to which a given projector \$E\$ belongs.

Here we derive independently the equivalence classes for pure states.\\

{\bf Theorem 3.} Let \$\ket{\psi}\$ be an arbitrary state vector. Any equivalence class \$[E]\$, i. e., the set of all projectors that are twins in \$\ket{\psi}\$ with an arbitrary given projector \$E\$, consists of all projectors of the form $$E_0+\bar E,\eqno{(6a)}$$ where $$E_0\equiv \Big(E\ket{\psi}\Big/||E\ket{\psi}||\Big)\Big( \bra{\psi}E\Big/||E\ket{\psi}||\Big),\enskip
\mbox{if}\enskip E\ket{\psi}\not=0;\eqno{(6b)}$$ and $$E_0\equiv 0\quad \mbox{if}\quad E\ket{\psi}=0.\eqno{(6c)}$$ By
\$\bar E\$ is denoted any projector orthogonal to both
\$E_0\$ and \$\ket{\psi}\bra{\psi}\$.\\

{\bf Proof} of Theorem 3. First we establish that \$E_0\in [E]\$. Substituting \$E_0\$ from (6b) in \$E_0\ket{\psi}\$, we obtain $$E_0\ket{\psi}=\Big(E\ket{\psi}\Big/||E\ket{\psi}||\Big) \Big(\bra{\psi}E\Big/||E\ket{\psi}||\Big)\ket{\psi}=E\ket{\psi},$$
because $$\bra{\psi}E\ket{\psi}=||E\ket{\psi}||^2.$$

Further, it is obvious from (6b) that \$EE_0=E_0\$, i. e., that \$E_0\$ is a subprojector of \$E\$: \$E_0\leq E\$. It is also obvious from (6b) that \$E\$ can be replaced with any other element of the class \$[E]\$ in the very definition of \$E_0\$. Hence, \$E_0\$ is a sub-projector of every projector in \$[E]\$. It is the minimal element in the class.

Since Theorem 3 requires \$\bar E\$ to be orthogonal to
\$\ket{\psi}\bra{\psi}\$, it follows that \$\bar E\ket{\psi}=0\$ and \$(E_0+\bar E)\ket{\psi}= E_0\ket{\psi}\$. Thus, all projectors given by (6a) belong to the equivalence class \$[E]\$.

Next we assume that \$E'\in [E]\$. Since it has been proved
that \$E_0\in [E]\$, and that it is the minimal element of
the class, the operator \$E'-E_0\$ is a projector orthogonal
to \$E_0\$. Further, \$(E'-E_0)\ket{\psi}=0\$  implies that
\$(E'-E_0)\$ is orthogonal to \$\ket{\psi}\bra{\psi}\$. Hence, \$E'\$ is of the form (6a) with (6b), and the claims of Theorem 2 are valid.

The case \$E\ket{\psi}=0\$ is proved analogously.\hfill $\Box$\\

One should notice that \$E\ket{\psi}\Big/||E\ket{\psi}||\$ is the state into which \$\ket{\psi}\$ changes when \$E\$ occurs  ideally in
\$\ket{\psi}\$ (cf Definition 1). (The former state was called the L\"{u}ders state in \cite{FHAMP}.)\\

Now we go to more general observables, but we confine ourselves to those with a purely discrete spectrum (allowing it to be infinite). The next definition is so formulated that the text of its part A can be utilized again below.\\

{\bf Definition 3.} Let \$\ket{\psi}\$ be a state vector, \$O=\sum_ko_kE_k,\enskip k\not= k'\enskip\Rightarrow
\enskip o_k\not= o_{k'}\$ and
\$O'=\sum_lo_lF_l,\enskip l\not= l'\enskip\Rightarrow\enskip o_l\not= o_{l'}\$ be two observables in spectral form. They are {\bf twin observables} or observables that are twins
in \$\ket{\psi}\$ if

{\bf A)} In both observables
{\bf all}  positive-probability
eigenvalues can be renumerated by an index \$m\$ {\bf common to the
two observables} so that all corresponding eigen-projectors
\$E_m\$ and \$F_m\$  satisfy a certain condition.

{\bf B)} For each value of \$m\$, the corresponding eigen-projectors \$E_m\$
and \$F_m\$ satisfy the condition that
they are twin projectors:
$$\forall m:\quad E_m\ket{\psi}=F_m\ket{\psi}.\eqno{(7)}$$\\

Definition 3 generalizes to not-necessarily-composite state
vectors the definition given in \cite{FHPR}. In the special
case when the two observables themselves act equally on the state vector, as in the illustration given in the Introduction (the singlet state), the twin observables were called algebraic ones, a special case of the physical ones.

Now we give two alternative definitions of twin observables.\\

{\bf Corollary 1.} Two {\bf observables} \$O=\sum_ko_kE_k,
\enskip k\not= k'\enskip\Rightarrow\enskip o_k\not= o_{k'}\$
and \$O'=\sum_lo_lF_l,\enskip l\not= l'\enskip\Rightarrow
\enskip o_l\not= o_{l'}\$ that are given in spectral form
are {\bf twins} in a pure state \$\ket{\psi}\$ if any one of the
following two conditions (i) and (ii) is valid, and only if both are
satisfied. The (composite, necessary and sufficient) conditions are:

{\bf (i) (a)} Condition A from Definition 3 is valid.

{\bf (i) (b)} All corresponding eigen-events \$E_m\$ and \$F_m\$ have equal probabilities in the given state: $$\forall m:\quad
\bra{\psi}E_m\ket{\psi}=\bra{\psi}F_m\ket{\psi}.\eqno{(8)}$$

{\bf (i) (c)} Ideal occurrence of the corresponding eigen-events gives the same state: $$\forall m:\quad
E_m\ket{\psi}\Big/||E_m\ket{\psi}||=
 F_m\ket{\psi}\Big/||F_m\ket{\psi}||.\eqno{(9)}$$

 {\bf (ii) (a)} Condition A from Definition 3 is valid.

{\bf (ii) (b)} If any of the eigen-events \$E_m\$ occurs
ideally in \$\ket{\psi}\$, then the corresponding eigen-event
\$F_m\$ of the second observable becomes certain in
the state that comes about, i. e., one has
$$\forall m:\quad E_m\ket{\psi}\Big/||E_m\ket{\psi}||=
F_m\Big( E_m\ket{\psi}\Big/||E_m\ket{\psi}||\Big)
\eqno{(10)}$$ (cf the Lemma).\\

{\bf Proof} follows from Theorems 1 and 2 in a straightforward way except for the sufficiency of condition (ii), which we prove as follows.

Definition 3 implies that \$\ket{\psi}=
\sum_{m'}E_{m'}\ket{\psi}\$. Further,
repeated use of condition (ii)(b), orthogonality and idempotency of the
\$F_m\$ projectors enables one to write $$\forall m:\quad F_m\ket{\psi}=
\sum_{m'}||E_{m'}\ket{\psi}||\times F_m
\Big[\Big( E_{m'}\ket{\psi}\Big/ ||E_{m'}
\ket{\psi}||\Big)\Big]=$$
$$\sum_{m'}||E_{m'}\ket{\psi}||
\times F_mF_{m'}\Big[\Big( E_{m'}\ket{\psi}
 \Big/ ||E_{m'}\ket{\psi}||\Big)
\Big]=$$ $$||E_m\ket{\psi}||\times F_m
\Big[\Big( E_m\ket{\psi}\Big/ ||E_m\ket{\psi}||\Big)
\Big]=$$ $$
||E_m\ket{\psi}||\times\Big[\Big( E_m\ket{\psi}
\Big/ ||E_m\ket{\psi}||\Big)
\Big]=E_m\ket{\psi}.$$ \hfill $\Box$\\

\section{SECOND GENERALIZATION: DELAYED TWIN EVENTS}

The first generalization of the notion of twin events and twin observables (in the preceding section) targeted the states of all \Q systems, not just the bipartite ones as treated in previous work. But it all applied to a \Q pure state given at one fixed moment. In the context of delayed twins that are going to be introduced, the twins from the preceding section will be called {\bf simultaneous twins}. Now we generalize further, allowing for unitary evolution separating the two twins and considering two moments \$t\geq t_0\$.\\

Henceforth we consider an arbitrary given pure state \$\ket{\psi ,t_0}\$ at an arbitrary given moment \$t_0\$. The \Q system is assumed to be isolated from its surroundings in a time interval \$[t_0,t]\$, \$t\geq t_0\$, so that  the change of state is governed by a unitary evolution operator \$U(t-t_0)\$, which we write shortly as \$U\$. The delayed state will be interchangeably written as \$U\ket{\psi ,t_0}\$ or as \$\ket{\psi ,t}\$.\\

{\bf  Definition 4.} Two events \$E\$ and \$F\$ are {\bf delayed twins} in the state \$\ket{\psi ,t_0}\$ for the time interval or delay \$(t-t_0)\$ if $$UE\ket{\psi ,t_0}=FU\ket{\psi ,t_0}\eqno{(11)}$$ is valid.\\

{\bf Remark 3.} It follows from the definition given by (11) that \$E\$ and \$UEU^{\dag}\$ are delayed twins. They are {\bf the trivial delayed twins}.\\

{\bf Proposition 1.} Let \$E\$ be an event. The equivalence class \$[UEU^{-1}]\$ of all events that are simultaneous twins with
\$UEU^{-1}\$ in \$\ket{\psi ,t}\$ consists of all delayed events from the equivalence class \$[E]\$ of all simultaneous twins of \$E\$ in the state \$\ket{\psi ,t_0}\$, i. e.,
$$[UEU^{-1}]=\{UE'U^{-1}:E'\in [E]\}. \eqno{(12)}$$\\

{\it Proof.} If \$E'\in [E]\$, then \$E'
\ket{\psi ,t_0}=E\ket{\psi ,t_0}\$ implying \$(UE'U^{-1})(U\ket{\psi ,t_0})=(UEU^{-1})(U\ket{\psi ,t_0})\$. Hence the set on the rhs of (12) is a subset of its lhs. On the other hand, if
\$F\in [UEU^{-1}]\$, then \$F(U\ket{\psi ,t_0})=(UEU^{-1})(U\ket{\psi ,t_0})\$. Multiplying this from the left with \$U^{-1}\$, one obtains \$(U^{-1}FU)\in
[E]\$. Since \$F=U(U^{-1}FU)U^{-1}\$, the lhs of (12) is a subset of its rhs. Hence, the two sides of the claimed relation are equal.\hfill $\Box$\\

{\bf Remark 4.} In intuitive terms, one might say that the evolution or delay splits the equivalence class \$[E]\$ into two equal copies; one remains unchanged as \$[E]\$ at \$t_0\$, and the other, the delayed clone, appears as \$[UEU^{-1}]\$ at \$t\$.\\

{\bf Remark 5.} Proposition 1 implies that the unitary evolution operator \$U\$ from \$t_0\$ till \$t\enskip\Big(t\geq t_0\Big)\$ maps all equivalence classes into which the set of all projectors (quantum logic) \$\cP(\cH)\$ is broken up with respect to \$\ket{\psi ,t_0}\$ into those regarding \$\ket{\psi ,t}\$.\\

{\bf Proposition 2.} Let \$E,E',F,F'\$ be four events.

{\bf A)} If \$E,F\$ are a pair of delayed twins in a state \$\ket{\psi ,t_0}\$ for the delay \$(t-t_0)\$, then so are \$E,F'\$  \IF \$F\$ and \$F'\$ are {\bf simultaneous twins} in \$\ket{\psi ,t}\$.

{\bf B)} If the events \$E,F\$ are  delayed twins in \$\ket{\psi ,t_0}\$ for the delay \$(t-t_0)\$, then so are \$E',F\$  \IF \$E\$ and \$E'\$ are {\bf simultaneous twins} in \$\ket{\psi ,t_0}\$.\\

{\bf Proof. A) Necessity.} Utilizing relation (11), we have \$UE\ket{\psi ,t_0}=F\ket{\psi ,t}\$ and \$UE\ket{\psi ,t_0}=F'\ket{\psi ,t}\$. Hence, \$F\ket{\psi ,t}=F'\ket{\psi ,t}\$, i. e., \$F\$ and \$F'\$ are simultaneous twins.

{\bf Sufficiency.} The relations \$UE\ket{\psi ,t_0}=
F\ket{\psi ,t}\$ and \$F\ket{\psi ,t}=F'\ket{\psi ,t}\$ imply \$UE\ket{\psi ,t_0}=F'\ket{\psi ,t}\$

{\bf B) Necessity.}  Assuming \$UE\ket{\psi ,t_0}=F\ket{\psi ,t}\$ and \$UE'\ket{\psi ,t_0}=F\ket{\psi ,t}\$, one obtains \$E\ket{\psi ,t_0}=E'\ket{\psi ,t_0}\$.

{\bf Sufficiency.} The relations \$UE\ket{\psi ,t_0}=
F\ket{\psi ,t}\$ and \$E\ket{\psi ,t_0}=E'\ket{\psi ,t_0}\$ imply \$UE'\ket{\psi ,t_0}=F.\ket{\psi ,t}\$.\hfill $\Box$\\

Note that in \P 2 the equivalence classes \$[E]\$ and \$[UEU^{-1}]\$ play symmetrical roles with respect to delay and anti-delay or inverse delay.\\

{\bf Theorem 4. A)} Let \$E\$ be an arbitrary event and let \$E'\$ be an arbitrary event from the equivalence class \$[E]\$ of all simultaneous twins with \$E\$ in \$\ket{\psi ,t_0}\$. Let \$F\$  also be an event. The events\$E'\$ and \$F\$ are delayed twins for a given time interval \$(t-t_0)\$ \IF \$F\$ is a simultaneous twin with the event \$UEU^{\dag}\$  in \$\ket{\psi ,t}\enskip\Big(\equiv U\ket{\psi ,t_0}\Big)\$.

{\bf B)}  Symmetrically: Let \$F\$ be an arbitrary event and \$F'\$  an arbitrary event from the equivalence class \$[F]\$ of all simultaneous twins with \$F\$ in \$\ket{\psi ,t}\$. Let \$E\$  also be an event. The events\$E\$ and \$F'\$ are delayed twins for \$(t-t_0)\$ \IF \$E\$ is a simultaneous twin with the event \$U^{\dag}FU\$  in \$\ket{\psi ,t_0}\$.\\

{\it Proof. A)} It follows from \P 2(A) that\$E'\$ and \$UEU^{-1}\$  are delayed twins. The former is a delayed twin with \$UE'U^{-1}\$, and this is, in turn, a simultaneous twin with \$UEU^{-1}\$ in the later state. Then \P 2(B) implies that \$E'\$ and \$F\$ are delayed twins \IF the latter is a simultaneous twin with \$UE'U^{-1}\$, and, due to transitivity, \IF \$F\$ is a simultaneous twin with \$UEU^{-1}\$ in the later state.

{\it B)} is proved analogously.\hfill $\Box\$\\

Note that also in \T 4 the equivalence classes \$[E]\$ and \$[UEU^{-1}]\$ play symmetrical roles with respect to delay and antidelay or inverse delay.\\

{\bf Definition 5.} {\bf Two pairs of delayed twin events} \$E,F\$ and \$E',F'\$ in the same pure state \$\ket{\psi ,t_0}\$ and for the same delay \$(t-t_0)\$ are {\bf equivalent} if \$E\$ and \$E'\$ are simultaneous twins in the earlier state \$\ket{\psi ,t_0}\$, and so are \$F\$ and \$F'\$ in the later state \$\ket{\psi ,t}\$.\\

Two Remarks, obvious consequences of \D 5, and an additional Remark help to fully understand the concept of equivalent pairs of delayed twin events.\\

{\bf Remark 6.} Two pairs of delayed twin events} \$E,F\$ and \$E',F'\$ in the same state \$\ket{\psi ,t_0}\$ for the same time interval \$(t-t_0)\$ are {\bf inequivalent} \IF {\bf both} \$E\$ and \$E'\$ are {\bf not} simultaneous twins in \$\ket{\psi ,t_0}\$ and \$F\$ and \$F'\$ are are {\bf not} simultaneous twins in \$\ket{\psi ,t}\$.\\

{\bf Remark 7.} It cannot happen that two pairs of delayed twin events \$E,F\$ and \$E',F'\$ are inequivalent with respect to \$\ket{\psi ,t_0}\$ and for \$(t-t_0)\$  because \$E\$ and \$E'\$ fail to be simultaneous twins in \$\ket{\psi ,t_0}\$ though \$F\$ and \$F'\$ are simultaneous twins in \$\ket{\psi ,t}\$ and symmetrically with respect to delay-antidelay.\\

{\bf Remark 8.} The claims of \T 4, \C 2 and \R 6 amount to saying that there is no cross-twinning, i. e., twinning across the equivalence classes. This is understandable intuitively in view of \R 4 and \R 5.\\

Now, following the claims of \T 2 from the preceding section, we give four necessary and sufficient conditions, or equivalent definitions, for {\bf delayed twin events}, valid when the given projector \$E\$ does not nullify the given state \$\ket{\psi ,t_0}\$.\\

{\bf Theorem 5.} Let \$E\ket{\psi ,t_0}\not= 0\$. Two events \$E\$ and \$F\$ are {\bf delayed twins} in  \$\ket{\psi ,t_0}\$ for \$(t-t_0)\$ (cf Definition 4) if any of the following four conditions is valid, and only if all four are. The (simple or composite) conditions (i)-(iv) read:

{\bf (i)} The opposite events \$E^c\$ and \$F^c\$ are delayed twins in \$\ket{\psi ,t_0}\$: $$UE^c\ket{\psi ,t_0}=F^cU\ket{\psi ,t_0}.\eqno{(13)}$$

\vspace{3mm}

{\bf (ii) (a)} The event \$E\$ has the {\bf same probability} in the state \$\ket{\psi ,t_0}\$ as the event \$F\$ in the delayed state \$\ket{\psi ,t}\$: $$\bra{\psi ,t_0}E\ket{\psi ,t_0}=\bra{\psi ,t}F\ket{\psi ,t}.\eqno{(14a)}$$

{\bf (ii) (b)} The changed state due to ideal occurrence of \$E\$ in the state \$\ket{\psi ,t_0}\$ is, after evolution, equal to the changed state brought about by ideal occurrence of the event \$F\$ in the delayed state \$\ket{\psi ,t}\$: $$U\Big(E\ket{\psi ,t_0}\Big/||E\ket{\psi ,t_0}||\Big)=F\ket{\psi ,t}\Big/||F\ket{\psi ,t}|| \eqno{(14b)}.$$ (Intuitively put: Collapse and evolution commute.)\\

{\bf (iii) (a)} When the state
\$E\ket{\psi ,t_0}\Big/||E\ket{\psi ,t_0}||\$, which is a component in the coherent state decomposition like (1), is delayed by \$U\enskip\Big(\equiv U(t-t_0)\Big)\$, \$F\$ is a certain event in the delayed state: $$U\Big(
E\ket{\psi ,t_0}\Big/ ||E\ket{\psi ,t_0}||\Big)=F\Big[U\Big(E\ket{\psi ,t_0}\Big/ ||E\ket{\psi ,t_0}||\Big)\Big].\eqno{(15a)}$$

{\bf (iii) (b)} When the state \$F\ket{\psi ,t}\Big/||F\ket{\psi ,t}||\$, which is a component in the corresponding coherent decomposition of
the state \$\ket{\psi ,t}\$, is inversely delayed, then the event \$E\$ is {\bf certain} in the past, inversely delayed state  \$U^{-1}\Big(F\ket{\psi ,t}\Big/||F\ket{\psi ,t}||\Big)\$:
$$U^{-1}\Big(F\ket{\psi ,t}\Big/||F\ket{\psi ,t}||\Big)=
E\Big[U^{-1}\Big(F\ket{\psi ,t}\Big/||F\ket{\psi ,t}||\Big)
\Big].\eqno{(15b)}$$

\vspace{3mm}

{\bf (iv) (a)} If the event \$E\$ occurs ideally in the state \$\ket{\psi ,t_0}\$, then the event \$F\$ is {\bf certain} in its corresponding delayed state \$U\Big(E\ket{\psi ,t_0}\Big/
||E\ket{\psi ,t_0}||\Big)\$.\

{\bf (iv) (b)} The analogous statement is valid for the opposite events \$E^c\$ and \$F^c\$ respectively.\\

{\bf Proof} consists in the simple fact that each of the four conditions is valid \IF the corresponding condition in Theorem 1 is valid for the events \$UEU^{\dag}\$ and \$F\$,
making them simultaneous twins in the state \$\ket{\psi ,t}\enskip\Big(\equiv U\ket{\psi ,t_0}\Big)\$. To make this transparent, we write
out the details item by item.

(i) It is straightforward to see that relation  (12) can be rewritten as $$\Big(UE^cU^{\dag}\Big)\ket{\psi ,t}=F^c\ket{\psi ,t}$$ (cf (2) in Theorem 1).

(ii) Relation (14a) can be written in the form $$\bra{\psi ,t}\Big(UEU^{\dag}\Big)\ket{\psi ,t}=
\bra{\psi ,t}F\ket{\psi ,t}.$$ Further, relation (14b) is equivalent to $$\Big(UEU^{\dag}\Big)\ket{\psi ,t}\Big/||(UEU^{\dag})\ket{\psi ,t}||\Big)=F\ket{\psi ,t}\Big/||F\ket{\psi ,t}||$$ (cf Theorem 1 (ii)).

(iii) Relation (15a) can take the form $$\Big(UEU^{-1}\Big)\ket{\psi ,t}\Big/ ||(UEU^{-1})\ket{\psi ,t}||=F\Big[\Big(UEU^{-1}\Big)\ket{\psi ,t}\Big/ ||(UEU^{-1})\ket{\psi ,t}||\Big]$$ (cf Theorem 1 (iii) (a)).
Further, relation (15b) is equivalent to $$F\ket{\psi ,t}\Big/||F\ket{\psi ,t}||=\Big(UEU^{\dag}\Big)\Big(
F\ket{\psi ,t}\Big/||F\ket{\psi ,t}||\Big)$$ (cf (iii)(b) in Theorem 1).

(iv) Claim (a) is proved above in (iii) (a). Claim (b)
is proved analogously.\hfill $\Box$\\

\section{TWIN OBSERVABLES}

Now we go to  observables  with a purely discrete spectrum
(allowing it to be infinite). The next  definition is written in a redundant way, just like the analogous definition in the preceding section, in order to make possible repeated use of its condition A.\\

{\bf Definition 6.} Let
\$O=\sum_ko_kE_k,\enskip k\not= k'\enskip\Rightarrow
\enskip o_k\not= o_{k'}\$ and
\$O'=\sum_lo_lF_l,\enskip l\not= l'\enskip\Rightarrow
\enskip o_l\not= o_{l'}\$ be two observables in spectral form,
and \$\ket{\psi ,t_0}\$ a state vector. One is dealing with
{\bf delayed twin observables} or observables that are
delayed twins in the state
\$\ket{\psi ,t_0}\$ for the time interval or delay \$(t-t_0)\$ if

{\bf A)} {\bf All} of the
positive-probability  eigenvalues of \$O\$ and \$O'\$ in \$\ket{\psi ,t_0}\$ 
can be renumerated by a common index \$m\$ so that the
corresponding eigen-projectors \$E_m\$ and \$F_m\$ satisfy a given condition.

{\bf B)} Assuming validity of condition A, the corresponding eigen-projectors \$E_m\$ and \$F_m\$ satisfy the condition that they are
delayed twin projectors for each value of \$m\$: $$\forall m:
\quad UE_m\ket{\psi ,t_0}=F_m\ket{\psi ,t}, \eqno{(16)}$$
where \$U\equiv U(t-t_0)\$.\\

Next we give two alternative definitions of twin observables, which are potentially important for applications.\\

{\bf Theorem 6.} {\bf Two observables}
\$O=\sum_ko_kE_k,\enskip k\not= k'\enskip\Rightarrow
\enskip o_k\not= o_{k'}\$ and
\$O'=\sum_lo_lF_l,\enskip l\not= l'
\enskip\Rightarrow\enskip o_l\not= o_{l'}\$
(given in spectral form) are {\bf delayed twins}
in a pure state \$\ket{\psi ,t_0}\$ for \$(t-t_0)\$ if any one of the following two (composite) conditions (i) and (ii) is valid, and  only if both are satisfied. The conditions are:

{\bf (i) (a)} Condition A in \D 6 is satisfied.

{\bf (i) (b)} The corresponding eigen-events have equal probabilities in the respective states: $$\forall m:\quad \bra{\psi ,t_0}E_m\ket{
\psi ,t_0}=\bra{\psi ,t}F_m
\ket{\psi ,t}. \eqno{(17)}$$

{\bf (i) (c)} ideal occurrence of the
corresponding eigen-events gives the same state
up to the delay: $$\forall m:\quad U\Big(E_m\ket{\psi ,t_0}
\Big/||E_m\ket{\psi ,t_0}||\Big)= F_m\ket{\psi ,t}
\Big/||F_m\ket{\psi ,t}||.\eqno{(18)}$$

\vspace{3mm}

{\bf (ii) (a)} Condition A in Definition 6 is valid.

{\bf (ii) (b)} If any of the eigen-events \$E_m\$ occurs in
ideal \M in \$\ket{\psi ,t_0}\$, then the corresponding
 eigen-event \$F_m\$ of the second observable becomes
certain  in the state that comes about in the \M and
becomes delayed: $$\forall  m:\quad U\Big(E_m\ket{\psi ,t_0}
\Big/
||E_m\ket{\psi ,t_0}||\Big)=$$ $$
F_m\Big[U\Big( E_m\ket{\psi ,t_0}\Big/||E_m\ket{\psi
,t_0}||\Big)\Big].\eqno{(19)}$$\\

{\bf Proof}  {\bf (i)} This claim coincides with that of Theorem 4, condition (ii). Hence, it has already been proved that this condition is necessary and sufficient for the observables being delayed twins.

{\bf (ii). Necessity} is obvious from Theorem 4 (iv).

{\bf Sufficiency.} Repeated use of condition (ii), orthogonality and idempotency of the
\$F_m\$ projectors, and the fact that
$$\sum_{m'}E_{m'}\ket{\psi ,t_0}=\ket{\psi ,t_0},$$ enable one to write $$\forall m:\quad F_mU\Big[
\ket{\psi ,t_0}\Big]=
\sum_{m'}||E_{m'}\ket{\psi ,t_0}||\times F_m
U\Big[\Big( E_{m'}\ket{\psi ,t_0}\Big/ ||E_{m'}
\ket{\psi ,t_0}||\Big)\Big]=$$
$$\sum_{m'}||E_{m'}\ket{\psi ,t_0}||
\times F_mF_{m'}U\Big[\Big( E_{m'}\ket{\psi ,t_0}
 \Big/ ||E_{m'}\ket{\psi ,t_0}||\Big)
\Big]=$$ $$||E_m\ket{\psi ,t_0}||\times F_mU\Big[\Big( E_m\ket{\psi}\Big/ ||E_m\ket{\psi ,t_0}||\Big)
\Big]=$$ $$
||E_m\ket{\psi ,t_0}||\times U\Big[\Big( E_m\ket{\psi ,t_0}
\Big/ ||E_m\ket{\psi ,t_0}||\Big)
\Big]=UE_m\ket{\psi ,t_0}.$$ \hfill $\Box$\\

Claim (i) (c) of Theorem 6, which is the corner stone of the next theorem, can be put intuitively and roughly as follows: - Selective collapse and evolution commute for delayed twin observables.-\\

Theorem 6 (i) has an obvious consequence that may be very important for applications (see e. g. subsection 5.2 below). Hence it must be spelled out though its precise formulation is cumbersome. To emphasize its expected importance, we write it as a theorem. In intuitive and imprecise but concise terms we can put its claim as: - Nonselective collapse commutes with evolution for delayed twin observables.-\\

{\bf Theorem 7.} Let two observables \$O\$ and \$O'\$ be delayed twins
for a given state \$\ket{\psi ,t_0}\$ and for a given interval \$(t-t_0)\$ as in Theorem 6. We compare two situations.

{\bf A)} Nonselective ideal \M of \$O\$ , i. e. its ideal \M on an entire ensemble of \Q systems, in the state \$\ket{\psi ,t_0}\$ is performed at \$t_0\$ (thus all positive-probability results appear on some systems in the ensemble). The \M converts the pure state into the {\bf mixture} $$\rho_{t_0}=\sum_mw_m^0
\rho_m^0,\eqno{(20a)}$$ $$\forall m:\enskip \rho_m^0
\equiv  \Big(E_m\ket{\psi ,t_0}\Big/||E_m\ket{\psi
,t_0}||\Big)\Big( \bra{\psi ,t_0}E_m\Big/||E_m\ket{\psi
,t_0}||\Big),\eqno{(20b)}$$ $$\forall m:\quad w_m^0\equiv\bra{\psi ,t_0}E_m\ket{\psi ,t_0}.\eqno{(20c)}$$

{\bf B)} As an alternative, we take the situation when the pure state \$\ket{\psi ,t_0}\$ evolves unitarily till the moment \$t\$, becoming \$\ket{\psi ,t}\equiv U\ket{\psi ,t_0}\$, and then ideal nonselective \M of the {\bf delayed twin observable} \$O'\$ is carried out on the ensemble described by \$\ket{\psi ,t}\$ resulting in the mixed state $$\rho_t\equiv\sum_mw_m^t
\rho_m^t,\eqno{(21a)}$$ $$\forall m:\enskip \rho_m^t
\equiv  \Big(F_m\ket{\psi ,t}\Big/||F_m\ket{\psi ,t}||\Big)
\Big(\bra{\psi ,t}F_m\Big/ ||F_m\ket{\psi ,t}||\Big),\eqno{(21b)}$$ $$\forall m:\quad w_m^t\equiv\bra{\psi ,t}F_m\ket{\psi ,t}.\eqno{(21c)}$$

It is {\bf claimed} that:

{\bf C)} The unitarily evolved mixed state \$U\rho_{t_0}U^{-1}\$ and the state \$\rho_t\$ are {\bf equal}: $$U\rho_{t_0}U^{-1}=\rho_t.\eqno{(22)}$$

{\bf D)} The statistical weights \$w_m^0\$ of the pure states \$\rho_m^0\$, which are also the probabilities of the individual results \$o_m\$ in the \M of \$O\$ in \$\ket{\psi ,t_0}\$, are {\bf equal} to the statistical weights \$w_m^t\$
of the corresponding pure states \$\rho_m^t\$, which are also the probabilities of the results \$o'_m\$ of the \M of the delayed twin observable\$O'\$ in \$\ket{\psi ,t}\$: $$\forall m:\quad\bra{\psi ,t_0}E_m \ket{\psi ,t_0}=w_m^0\quad\mbox{\bf =}\quad w_m^t=
\bra{\psi ,t}F_m\ket{\psi ,t}.\eqno{(23)}$$\\

{\bf Corollary 2.} Delayed twin observables have the {\bf chaining property} in the following sense. If
two observables
\$O=\sum_ko_kE_k,\enskip k\not= k'\enskip\Rightarrow
\enskip o_k\not= o_{k'}\$ and
\$O'=\sum_lo_lF_l,\enskip l\not= l'
\enskip\Rightarrow\enskip o_l\not= o_{l'}\$ (given in spectral form) are delayed twins in the pure state \$\ket{\psi ,t_0}\$ for the time interval \$(t-t_0)\$, and if the latter observable and a third observable
\$O''=\sum_no''_nE_n,\enskip n\not= n'\enskip\Rightarrow\enskip o''_n\not= o''_{n'}\$ are delayed twins in the state \$\ket{\psi ,t}\enskip\Big(\equiv U(t-t_0)\ket{\psi ,t_0}\Big)\$ for \$\Big((t+t')-t\Big)\$, then the first observable and the third one are delayed twins in the state \$\ket{\psi ,t_0}\$ for \$\Big((t+t')-t_0\Big)\$.

{\bf Proof} is obvious.\\

{\bf Remark 9.} The notion of a chain of delayed twin observables comes close to the well known von Neumann chain (see e. g. \cite{vNchain}) and the infinite regress to which it leads (cf e. g. \cite{Hartle}). Since the former is based on the concept of delayed twins, and they are well understood in terms of the preceding results, the former chain may perhaps contribute to a better understanding of the latter.\\

{\bf Remark 10.} The chaining property of delayed twins may come close also to the concept of Consistent Histories, which was extensively studied by Griffiths \cite{Griff}, Gell-Mann  and Hartle \cite{GMH1} and \cite{GMH2} (who called it 'decoherent histories'), as well as by Omnes \cite{O}. This concept has the ambition to give a new interpretation of \QM . (It will not be further discussed in this article.)\\

The delayed-twin concept can find many applications in \QM as will be shown in follow-ups. In the next and last section we sketch a few applications.\\

\section{APPLICATIONS}

Now we describe shortly a few cases where the delayed-twin-observables concept can be seen to appear. By this we do not care whether it is useful or not in the mentioned example.\\

\subsection{Some Preparations of  Quantum Experiments}

Let us take the concrete example of preparing a $1/2$-spin particle in spin-up state by the Stern-Gerlach measuring instrument adapted for preparation. To this purpose, there is open space instead of the upper part of the screen (or instead of an upper detector). Let event \$E\$ for the particle be "passage through the upper part" at \$t_0\$, and let event \$F\$ for it be  \M at \$t,\enskip t>t_0\$, to the right of the described preparator. Imagine that the geometry is such that \$F\$ occurs \IF so does \$E\$ in the previous moment (when the preparation took successfully place). The two events are delayed twins according to condition (or definition) (iv) in Theorem 5.

One should not be confused by the fact that the event \$E\$ does not actually 'occur' in the quantum-mechanical sense in the described preparation because it is not observed (measured). Theoretically one can assume that it does because ideal selective position \M of \$E\$ does not change the up-going component in the Stern-Gerlach instrument. It is, of course, simpler to view just the collapse caused by the ideal occurrence of \$F\$ at the later moment of actual \m . But, as Bohr would say, "visualization" (classical intuition) does not allow us to understand the occurrence of \$F\$ unless \$E\$ has occurred (though unobserved) at \$t_0\$.

Whatever kind of \M is performed at \$t\$, we can again insert position collapse to the spatial domain occupied by the measuring instrument immediately before the actual \M takes place because it does not change the component of the state of the particle at \$t\$, which (locally) interacts with the instrument.\\

\subsection{A Puzzle in Understanding the Real Scully et al. Experiment}

Scully et al. have performed a very sophisticatedly upgraded two-slit interference-or-which-way experiment
\cite{Scully}. Instead of 'passing two slits', two excited atoms undergo cascade de excitation coherently emitting a pair of photons in opposite directions. Detectors are placed to the right of the atoms to detect the right-going photon, and detectors are put to the left to detect the photon moving to the left. The experimental arrangement is such that the photon that moves to the right is detected before the left-moving one reaches a detector.

Without going into details of the intricate experiment, theoretical analysis of the experiment led to a puzzle. Namely, Scully et al. have shown
\cite{Scully2} that one can view the experiment so that the right-moving photon undergoes spatial collapse at (various positions of) the corresponding detector, then one evaluates the effect of this collapse on the left-moving photon on account of the entanglement between the two photons, and at the end one evaluates the relative frequencies of the coincidence detections.

This first-quantization description of Scully et al. is precisely what one would expect in the Copenhagen approach. The puzzle was due to the fact that the present author gave a successful  description of the same experiment \cite{Scully} without collapse of the right-moving photon \cite{FHScully}. Which of the two mutually contradicting pictures is the correct description of the experiment?

The puzzle is immediately solved by Theorem 7. If \$t_0\$ is the moment of collapse of the right-moving photon, observation of its detection at a certain location of the right-placed detector can be viewed as ideal measurement (by the sentient observer) of an observable \$O\$, and the corresponding observation of the same detection result at moment \$t\$, when the left-moving photon has reached a detector, can be viewed as ideal \M of a delayed twin observable \$O'\$ because a reading at \$t\$ can be done \IF the same result could (or could have been) done at \$t_0\$. Then the commutation of collapse and evolution, with equality of the probabilities, claimed in Theorem 7 explains the puzzle.

As my late teacher professor R. E. Peierls used to say, "paradoxes are useful because one can learn from their solutions". Here we learn that \Q experiments can be viewed not only in different, but also in different mutually exclusive ways. Each view provides us with some specific advantage.\\

\subsection{Exact Quantum Measurement}

Let the object on which the \M is performed be
subsystem A, and let the measuring instrument be subsystem B.
Further, let
\$O_A=\sum_ko_kE_A^k.\enskip k\not= k'\enskip\Rightarrow
\enskip o_k\not= o_{k'}\$ be the observable that is measured,
and let \$P_B=\sum_ko'_kF_B^k.\enskip k\not= k'\enskip
\Rightarrow\enskip o'_k\not= o'_{k'}\$ be the so-called
'pointer observable' (both given in spectral form). The
measuring instrument 'takes cognizance' of a result \$o_k\$
(or rather of the occurrence of the corresponding eigen-event
\$E_A^k\$) in terms of
the 'pointer position' $o'_k\$ (actually by the occurrence of \$F_B^k\$) that corresponds to it. (Note the common index.)

Exact \M (henceforth only '\m ') is defined in \QM by the
{\bf calibration condition} \cite{BLM}, \cite{FH}: If an event \$E_A^k\$ is certain
(has occurred) in the initial state \$\ket{\psi ,t_0}_A\$ of
the object, then the corresponding  event \$F_B^k\$ on the
measuring instrument is certain (occurs) in
the final composite-system state
\$\ket{\Psi ,t}_{AB}^f\equiv U_{AB}\Big(\ket{\psi ,t_0}_A
\ket{\psi ,t_0}_B^i\Big)\$ of the
object-plus-instrument system after the \M interaction has ended. Here \$U_{AB}\equiv U_{AB}(t_f-t_i)\$ is the unitary operator that contains the \M interaction; \$\ket{\psi ,t_0}_B^i\$ is the initial or 'trigger-ready' state of the measuring instrument; and the indices \$f\$ and \$i\$ refer to the final and the initial moments respectively.

In Bohrian collapse-interpretations of \QM one says that the operator \$U_{AB}\$ describes pre\M (\M short of collapse) \cite{BLM}; in no-collapse, relative-state interpretations it is the entire dynamical law of \m \cite{Everett1}, \cite{Everett2} . We shall write '(pre)\m ', to keep an open mind about the interpretation.

In algebraic form this reads
$$\ket{\psi ,t_0}_A=E_A^k\ket{\psi ,t_0}_A\quad
\Rightarrow\quad \ket{\Psi ,t}_{AB}^f=
F_B^k\ket{\Psi ,t}_{AB}^f\eqno{(24)}$$ (cf the Lemma).\\

An arbitrary initial state \$\ket{\psi ,t_0}_A\ket{\psi ,t_0}_B^0\$ can be decomposed into its sharp measured-observable eigenvalue components $$\ket{\psi ,t_0}_A
\ket{\psi ,t_0}_B^0=\sum_k||E_A^k\ket{\psi ,t_0}_A||\times\Big((\ket{\psi ,t_0}_A \Big/||E_A^k\ket{\psi ,t_0}_A||)
\ket{\psi ,t_0}_B^0\Big).$$ The components are precisely those states into which selective ideal \M of the individual values \$o_k\$ of the measured observable would turn the initial state. On the other hand, the calibration condition, and the fact that the evolution operator is linear, imply that \$U\$ would evolve each component into a final state with the corresponding sharp value of the pointer observable.

According to Theorem 6 condition (ii), the (pre)measurement evolution makes the measured observable and the pointer observable delayed twins in the initial state for the interval \$(t_f-t_i)\$.

At first glance one might jump to the conclusion that every case of opposite-subsystem delayed twin observables in a bipartite state can be viewed as \m . The fallacy is in
the fact that \m , being defined in terms of the calibration
condition, makes the measured observable and the pointer
observable twins in {\bf all} initial states of the measured object A. But still hopefully the delayed twin concept for opposite-subsystem
observables is the backbone of the notion of \M in \qm .

The measured observable and the pointer one , being twins,
'inherit' the nice properties of twins. No doubt, the most
important one is the so-called {\bf probability
 reproducibility condition} \cite{BLM}, \cite{FH} (cf Theorem 5 (i) (b) above). It says that the
 probability of a result \$o_k\$ in the initial state
 \$\ket{\psi ,t_0}_A\ket{\psi ,t_0}_B^0\$ {\bf equals}
 the probability of the corresponding pointer position
\$o'_k\$ in the final state: $$\forall k:\quad
\bra{\psi ,t_0}_AE_A^k \ket{\psi ,t_0}_A=
 \bra{\Psi ,t}_{AB}F_B^k\ket{\Psi ,t}_{AB}.
\eqno{(25)}$$

The probability reproducibility condition is crucial
for \M in ensembles, because it enables the statistical
information to become transferred from the ensemble of \Q objects
to the ensemble of results (relative frequencies) on the measuring instrument.\\

Foundationally-minded physicists prefer nondemolition
(synonyms: repeatable, first kind) \m , in which the value
\$o_k\$ of the measured observable, if sharp in the initial
state \$\ket{\psi ,t_0}_A\$, remains preserved (non demolished);
hence it can be checked in an immediately repeated \m .
Unfortunately, numerous laboratory \m s are of the opposite,
demolition type. (One should only be reminded of
experiments with photons, in which the photon becomes absorbed in a detector at the end.)

One wonders about the meaning of the result in demolition \m s
when the results cannot be checked. The delayed-twin concept
might be helpful here. The very definition  \$\forall k:
\Big(U_{AB}E_a^k\Big)\Big(\ket{\psi}_A\ket{\psi}_B^0\Big)=
F_B^kU_{AB}\Big(\ket{\psi}_A\ket{\psi}_B^0\Big)\$ of delayed twins
can be slightly changed into the form
$$\forall k:\quad   U_{AB}\Big[
\Big(E_a^k\ket{\psi}_A\Big/||E_A^k\ket{\psi}_A||\Big)
\ket{\psi}_B^0\Big]=F_B^k\ket{\Psi}_{AB}^f\Big/
||F_B^k\ket{\Psi}_{AB}^f||\eqno{(26)}$$ because
\$||E_A^k\ket{\psi}_A||=||F_B^k\ket{\Psi}_{AB}^f||\$
is the square root of the probability (cf (21) (b)).

Relation (26) tells us that the part of the final state
that corresponds to a sharp result \$o'_k\$ of the pointer observable  is actually
the time-evolved part of the initial state with a sharp
value of the measured observable.

In Copenhagen-inspired collapse interpretations of \qm , the usefulness of relation (26) is questionable because it suggests that collapse
of the final state to a definite result is tightly bound
with retroactive collapse at the initial moment \cite{FH}, \cite{FH2}.

The relation
is more useful in Everettian relative-state interpretations,
where  'collapse' is replaced by 'belonging to a branch'.
Then, what (26) says is that the entire process of \M with the
initial and the final moment belong to the same branch.

Even in nondemolition \M one has trouble answering the question
"Does \M create or find the result?" Relation (26) suggests
the latter answer in some sense.\\

Finally, the chaining property of delayed twin observables (cf Corollary 2) enables the \M results to be transferred to another system. For instance, an observable \$O_A\$ is measured in terms of a pointer observable \$P_B\$ (delayed twins). Then a human observer (not distinct from another measuring instrument quantum-mechanically) takes cognizance of the results in terms of his consciousness contents, which is, e. g., observable \$O'\$. This experimenter can communicate the results (I use the plural because I have an ensemble \M in mind) to another human being, or write them down etc. This would be another link in the chain in terms of a fourth observable that is a delayed twin with the third.

\end{document}